\documentclass[a4paper,11pt]{article}
\pdfoutput=1 

\usepackage{jcappub} 

\usepackage[T1]{fontenc} 
\usepackage{bm}
\usepackage{graphicx,subfig,color}
\usepackage{amssymb,amsmath}
\usepackage{slashed}
\usepackage{array}
\usepackage{hyperref}
\usepackage{orcidlink}
\usepackage{cleveref}

\title{\boldmath Effect of $f(R,T)$ theory of gravity on the properties of strange quark stars.}

\author[a,b,1]{Suman Pal\orcidlink{0009-0000-5944-4261},\note{Corresponding author}
}
\author[a,b]{Gargi Chaudhuri \orcidlink{0000-0002-8913-0658}}
\affiliation[a]{Physics Group, Variable Energy Cyclotron Centre, 1/AF Bidhan Nagar, Kolkata 700064, India}
\affiliation[b]{Homi Bhabha National Institute, Training School Complex, Anushakti Nagar, Mumbai 400085, India}

\emailAdd{sumanvecc@gmail.com}
\emailAdd{gargi@vecc.gov.in}

\abstract{ In this study, we investigate strange quark stars within the framework of modified $f(R,T)$ gravity, where $R$  represents the Ricci scalar and $T$ denotes the trace of the energy-momentum tensor, specifically defined as $ f(R,T) = R + 2\chi T $. The equation of state is obtained with the different forms of the MIT bag model and quark mass model with medium effects and self-consistent thermodynamical treatment. We find that negative values of $\chi$ significantly increase both the mass and radius of the quark star. The inclusion of $\chi$ helps to
satisfy recent the astrophysical constraints on the mass-radius relationship. We have also constrained  the values of  $\chi$ for each EoS, based on the observed maximum mass and corresponding radius, demonstrating that the inclusion of this parameter helps to address the challenges posed by both the GW190814 event and NICER observations of PSR J0030+0451. We  also observe that the inclusion of $f(R,T)$ gravity leads to an increase in both the maximum mass, by about $ (0.23- 0.27)~M_{\odot}$, and the corresponding radius, by approximately (1.5-2.0)~\text{km}, depending on the chosen equation of state.
}

\begin{document}
\maketitle
\flushbottom

\section{Introduction} \label{sect:intro} 
The investigation of neutron stars (NSs) serves as a valuable method for examining cold, dense matter and strongly interacting matter at high densities, conditions which are otherwise difficult to attain in Earth-based laboratories. In recent times, 
 there has been significant progress in availability of observational constraints which   makes the study of neutron stars  even  more challenging.  The estimate of the mass and radius of the central compact object HESS J1731-347 \cite{2022NatAs}: $M=0.77^{+0.20}_{-0.17}M_{\odot}$,~$R=10.4^{+0.86}_{-0.78}~km$, makes it the lightest neutron star known to date. It has been a potential candidate for a strange star. We also focus on the secondary companion of the remarkable GW190814 \cite{LIGOScientific:2020zkf} event, a compact object with a mass ranging from 2.50-2.67 $M_{\odot}$ and it has been challenging to explain the massive neutron star. In the simple MIT bag model without adding vector interaction,  it is difficult to satisfy the constraint of  ~$2M_{\odot}$~ limit from the observational constraints. One alternative approach to address this issue is to explore these objects within the framework of modified gravity theories.

The discovery of the universe's accelerated expansion and the existence of dark matter fundamentally challenge our understanding of gravity. Theories like General Relativity, while being highly successful in many contexts, struggle to explain these phenomena without introducing hypothetical entities like dark matter and dark energy, whose nature remains unknown.  It has been shown that cosmic acceleration can arise due to small corrections to the usual action of General Relativity and eliminates the need for a non zero cosmological constant or any form of dark energy \cite{Carroll:2003wy}. This led to the introduction of f(R) gravity which showed that it can account for the accelerated expansion of the Universe without the need of dark energy. f(R) theories which satisfy solar system tests and also unifies inflation with late time cosmic acceleration in the framework of a single model have been discussed in \cite{Nojiri:2007as}. The equation of motion for massive particles in a class of generalized gravitational models were studied \cite{Bertolami:2007gv} in which a coupling between matter Lagrangian and an arbitrary function f(R) of Ricci scalar R was considered. This gives rise to an extra force resulting in non geodesic equation of motion of the particles. This theory was applied in \cite{Poplawski:2006ey} where the cosmological constant  $\Lambda$ was considered to be a function of the trace of the energy-momentum tensor,  better known by $\Lambda (T)$ theory, a model for interacting dark energy. 

$f(R,T)$ modified theory of gravity which is another modification of the Einstein's general theory of relativity was first considered in \cite{Harko:2011kv}. Here the gravitational Lagrangian is a function of The Ricci Scalar($R$) and the trace of the energy-momentum tensor$(T)$. The reason for including $T$ is accounted for by quantum effects arising from bulk viscosity and other imperfections in the fluid. The choice of  functional form of $f$ is dictated by the nature of the matter source and this has been studied in details in the context of cosmology in \cite{Harko:2011kv}. $f(R,T)$ theory of gravity have been applied for  both cosmological \cite{Sharif:2012zzd,Shabani:2014xvi,Baffou:2015dna,Correa:2015qma} and astrophysical sources \cite{Hansraj:2018jzb,Pappas:2022gtt,Nashed:2023uvk,Pretel:2020oae,Pretel:2021kgl,Murshid:2023xsw,Batool:2024axw}. There are some studies on quark stars based on the  $f(R,T)$ theory. 

 First-principles methods are not applicable to describe quark matter at densities present inside the neutron star cores. This limitation arises due to the sign problem in lattice Monte Carlo simulations at non-zero chemical potentials \cite{deForcrand:2009zkb}, and the effectiveness of perturbative QCD is constrained to significantly higher densities \cite{Kurkela:2009gj}. Numerous efforts have been made to incorporate nonperturbative effects into increasingly sophisticated models, as perturbative QCD proves insufficient for addressing the equation of state (EoS) of quark matter. 
In recent studies, researchers have extensively employed phenomenological quark models like the MIT bag model  \cite{chodes1974,glendenning2012compact, Sen:2021cgl, Podder:2023dey,Weissenborn:2011qu}, vector bag model \cite{Cierniak:2019hhe,Lopes:2020btp,Kumar:2022byc,Pal:2023quk}and quark mass model or quasi-particle model \cite{Lugones:2022upj,Zhang:2021qhl, pen:23prc_qmdd,Ma:prd23sep,peng2001a, Chu_2014, Benvenuto95} to investigate the thermodynamic properties of strange quark matter, quark stars, and hybrid stars. These models typically account for all interactions among quarks through bag pressure or an equivalent quark mass.

In the past, the study of strange quark stars using the $f(R,T)$ theory of gravity have utilized the  MIT bag model with simple forms without using charge neutrality and chemical equilibrium conditions.
In this work, we study the strange quark star with realistic equations of state along with proper conservation laws. Our equation of state includes medium effects and incorporates a proper self-consistent thermodynamic treatment. In the realistic quark models, the equation of state parameters is constrained by incorporating the  Bodmer-Witten conjecture \cite{Torres:2012xv,Farhi:1984qu}. It tells that for 3 flavour quark matter, $\frac{\varepsilon}{\rho}<930 MeV$ at zero pressure where as for 2 flavour quark matter $\frac{\varepsilon}{\rho}>930 MeV$.  In this work, we use different versions of the  MIT bag model as well as the quark mass model. We incorporate the functional forms of the modified gravity as $f(R,T)=R+2\chi T$, the curvature part being the same as  Einstein's original theory. This results in the modification of the Tolman-Oppenheimer-Volkoff (TOV) equations \cite{Oppenheimer:1939ne,Tolman:1939jz} equation. The parameter $\chi=0$ corresponds to the original TOV equations. Using different values of $\chi$, we study the mass-radius diagram of the strange quark stars and also compare them with the astrophysical observations.

This paper is organized as follows: In Sec.~\ref{sec:formalism}, we outline the $f(R,T)$ theory of gravity and the equation of state for quark matter. In Sec.~\ref{sec:results}, we present the numerical results. Finally, we summarize our findings in Sec.~\ref{sec:conclusion}.

\section{Formalism} \label{sec:formalism}
\subsection{\textbf{$f(R,T)$ theory of gravity}}\label{sec:f(R,T)_theory}  
In this section, we provide a brief overview of the modified TOV equation, with detailed information being available in references \cite{Burikham:2016cwz,Carvalho:2017pgk,Carvalho:2020czc,Lobato:2020fxt,Moraes:2015uxq,Deb:2017rhc,Deb:2017rhd,Deb:2018sgt}.
The action of the $f(R,T)$ theory of gravity is proposed by \cite{Harko:2011kv}
\begin{equation}\label{eq:action}
  \mathcal{S}=\int d^{4}x\sqrt{-g}\left[  \frac{f(R,T)}{16\pi}+\mathcal{L}_{m}\right].
\end{equation} 
where $f(R,T)$ is an arbitrary function of the Ricci scalar $R$ and the trace of the energy-momentum tensor ($T = g^{\mu \nu} T_{\mu \nu}$). The factor $\sqrt{-g}$ is required to properly define the volume element in spacetime with the metric $g_{\mu\nu}$.
The matter Lagrangian density is denoted by $\mathcal{L}_{m} $, and the stress-energy tensor for matter is defined as
\begin{equation}\label{eq:def_T_mu_nu}
T_{\mu \nu }=-\frac{2}{\sqrt{-g}}
\frac{\delta \left( \sqrt{-g}L_\mathrm{m}\right) }{\delta g^{\mu \nu }}\, ,
\end{equation} 
By applying the principle of least action to the expression in \eqref{eq:action}, we obtain the resulting field equations.
\begin{multline}\label{eq:field_eqn}
  f_{R}(R,T)R_{\mu\nu}-\frac{1}{2}f(R,T)g_{\mu\nu}+(g_{\mu\nu}\Box-\nabla_{\mu}\nabla_{\nu})f_{R}(R,T)=8\pi T_{\mu\nu}-f_{T}(R,T)(T_{\mu\nu}+\Theta_{\mu\nu}),
\end{multline}
where $f_R=\frac{\partial f(R,T)}{\partial R}$ and $f_T(R,T)=\frac{\partial f(R,T)}{\partial T}$, $\Box=\frac{\partial_{\mu}(\sqrt{-g} g^{\mu\nu}\partial_{\nu})}{\sqrt{-g}}$, $\nabla_{\mu} \rightarrow \text{covariant deriavtive}$ and $\Theta_{\mu\nu}=g^{\alpha \beta} \frac{\delta T_{\alpha \beta }}{\delta g^{\mu \nu}}$. 
We will assume the energy-momentum tensor of a perfect fluid, i.e., $T_{\mu\nu}=(\varepsilon+p)u_\mu u_\nu-pg_{\mu\nu}$, with $\varepsilon$ and $p$ respectively representing the energy density and pressure of the fluid and $u_\mu$ being the four-velocity tensor.
In this study, we focus on the function \( f(R,T) = R + 2\chi T \). By substituting this form of \( f(R,T) \) into the field equations, the resulting modified Einstein equations are obtained. 
\begin{eqnarray}
&&G_{\mu\nu}=8\pi T_{\mu\nu}+\chi [Tg_{\mu\nu} +2(T_{\mu\nu}+pg_{\mu\nu})],\label{eqcampo2}
\\
&&\nabla^{\mu}T_{\mu\nu}=-\frac{2\chi}{8\pi+2\chi} \left[\nabla^{\mu}(pg_{\mu\nu})+\frac{1}{2}g_{\mu\nu}\nabla^{\mu}T\right],\label{divtensor2}
\end{eqnarray}
To obtain the modified hydrostatic equilibrium equation, we employ a spherically symmetric metric (details can be found in \cite{Harko:2011kv,Moraes:2015uxq,Deb:2017rhc,Deb:2017rhd,Deb:2018sgt}). As a result, the modified Tolman-Oppenheimer-Volkoff equations \cite{Moraes:2015uxq} are expressed as follows:

\begin{equation}\label{eq:modi_tov_final} 
\begin{aligned}
\frac{dm}{dr}=& 4\pi \varepsilon r^2 +\frac{\chi}{2}(3\varepsilon-p)r^2,\\
\frac{dp}{dr}=&-(p+\varepsilon) \frac{4\pi p r+\frac{m}{r^2}-\frac{\chi}{2}(\varepsilon-3p)r}{(1-\frac{2m}{r})(1+\frac{\chi}{8\pi+2\chi})(1-\frac{1}{C_s^2})}.\\
\end{aligned}
\end{equation} 
whre $C_s^2$ is the speed of sound ($\frac{dp}{d\varepsilon}$)


\subsection{\textbf{Quark equations of state}} \label{sec:quark_eos}
We focus on quark matter that might exist inside a neutron star.
For the strange quark matter, we consider u, d, and s quarks in chemical equilibrium along with the charge neutral phase. The governing equations are given as 
\begin{equation} 
\begin{aligned}
    &\mu_d=\mu_u+\mu_e=\mu_s~\text{: chemical equilibrium conditions } \\ 
    &\frac{2}{3}\rho_u-\frac{1}{3}\rho_d-\frac{1}{3}\rho_s-\rho_e=0 ~\text{: charge neutrality conditions}\\ 
    & \rho=\frac{1}{3}(\rho_u+\rho_d+\rho_s) ~\text{: baryon number conservation} \\
\end{aligned}
\end{equation} 
In this work, we have considered the recently studied different versions of the  MIT bag model \cite{pal:23prd2_bag} as well as the quark mass model \cite{Lugones:2022upj,Zhang:2021qhl,pen:23prc_qmdd,Ma:prd23sep}.
\subsubsection*{\textbf{MIT bag model }} 
In this study, we have explored various versions of the MIT bag model to describe quark matter.
In the MIT bag model, the medium effect is taken through medium dependence of bag pressure, the quark mass being assumed to be constant. The medium dependence of bag pressure depends on the 
to the choice of ensemble, which
solves the inconsistency problem from thermodynamics
point of view, details being  given in \cite{pal:23prd2_bag}. If the bag pressure depends on the chemical potential, the grand-canonical ensemble formalism is appropriate. On the other hand, when the bag pressure depends on the density, the canonical ensemble is more suitable.
The models can be briefly described as follows:

\subsubsection*{\textbf{Constant bag pressure (MIT bag)}} 
In the case of  MIT bag model equation of state bag pressure ($B$) is taken to be constant  and  medium effects and vector interactions  are not included. 
\subsubsection*{\textbf{Constant bag pressure with vector interactions(MIT Vbag)}}
In the vector bag model, bag pressure is again taken to be constant and medium effect is not included but vector interaction is included.
\subsubsection*{\textbf{Chemical potential dependent bag pressure ($B(\mu)$)}} 
In this case, we consider the bag pressure to depend on the chemical potential~\cite{pal:23prd2_bag}
\begin{equation}\label{eq:B_mu_expression}
    B(\mu)=B_{as}+(B_0-B_{as})e^{\left[-\beta_{\mu}(\frac{\mu}{\mu_0})^2\right]}
\end{equation}
$B_{as}$ is the parameter at the asymptotic chemical potential, $B_{\mu}$ attains the value  $B_0$ at zero chemical potential and $\beta_{\mu}$ is  the parameter controlling the decrease in the value of $B(\mu)$ with chemical potential.
The model parameters $B_{as}, B_0,\beta_{\mu}~\text{and}~\mu_0$ are determined by checking the Bodmer-Witten  stability criteria \cite{Farhi:1984qu,pal:23prd2_bag}. 
The thermodynamical potential for the quark matter with chemical potential-dependent bag pressure along  with vector interaction reads as 
\begin{equation} \label{eq:omega_mit_vbag}
\begin{aligned}
 \Omega=-\frac{1}{\pi^2} \sum_{f=u,d,s} \int_0^{k_{f}} \frac{k^4}{\sqrt{k^2 + m_f^2}} \, dk-\frac{1}{2}m_V^2V_0^2+B(\mu)+\Omega_e
\end{aligned}
\end{equation} 
where $V_0$ is the vector field, $m_V$ is the mass of the vector meson, $ g_V $ is the vector coupling constant, combining mass of the vector meson and coupling constant we define, $G_V=\left(\frac{g_V}{m_V}\right)^2$. $\Omega_e$ is electron thermodynamic potential.
To account for the role of the vector meson, the chemical potential of the quark gets modified.
\begin{equation}
\mu_f=\sqrt{k_f^2+m_f^2}+g_VV_0
\end{equation} 
The quark number densities are given by 
\begin{equation}\
    \rho_f=-\frac{\partial \Omega}{\partial \mu_f}=\frac{k_f^3}{\pi^2}-\frac{\partial B(\mu)}{\partial \mu_f}
\end{equation} 
The equation of motion of meson fields being obtained by 
\begin{equation}
\frac{\partial \Omega}{\partial V_0}=0
\end{equation} 
The equation of state is obtained as 
\begin{equation} \label{eq:eos_mit_b_mu}
    \begin{aligned}
       \text{pressure}\rightarrow P=&-\Omega \\
        \text{energy density}\rightarrow \varepsilon=&-P+\sum_{f=u,d,s,e}\mu_f\rho_f \\ 
    \end{aligned}
\end{equation}
\subsubsection*{\textbf{Density dependent bag pressure ($B(\rho)$)}} 
To describe the density-dependent bag pressure, we use the canonical ensemble, where all thermodynamic quantities are derived from the energy density at zero temperature.  The interactions and medium effects are taken care of by the bag pressure $B(\rho)$. The form of density-dependent bag pressure is given as 
\begin{equation}
    B(\rho)=B_{as}+(B_0-B_{as})e^{\left[-\beta_{\rho}(\frac{\rho}{\rho_0})^2\right]}
\end{equation}  
The energy density for quark matter with density-dependent bag pressure and vector interaction is given by
\begin{equation}\label{eq:free_en} 
    \varepsilon=\sum_{f=u,d,s}\frac{3}{\pi^2}\int_0^{k_f} k^2\sqrt{k^2+m_f^2}dk+\frac{1}{2}m_V^2V_0^2+B(\rho)+\frac{1}{\pi^2}\int_0^{k_e} k^2\sqrt{k^2+m_e^2}dk
\end{equation} 
where the medium effects are incorporated in bag pressure by making it density-dependent since density is the appropriate intensive parameter in the canonical ensemble.
The quark chemical potential is modified due to the density-dependent bag pressure and vector interactions.
\begin{equation}
    \mu_f=\sqrt{k_f^2+m_f^2}+g_VV_0+\frac{\partial B}{\partial \rho_f}
\end{equation} 
The pressure is obtained using the Euler realtion :
\begin{equation} \label{eq:pressure_euler}
    P=-\varepsilon+\sum_{f=u,d,s,e}\mu_f\rho_f
\end{equation} 

\subsection*{\textbf{Quark mass model ($m(\mu),m(\rho)$) }} 
In contrast to the MIT bag model, in the quark mass model, the medium effect of this model is taken through the quark mass, bag constant having a fixed value independent of density. In this work, we have employed two versions of the quark mass model: the first is the quark mass dependent on chemical potential, referred to as the quasi-particle model, and the second is the quark mass dependent on density, referred to as the quark mass density-dependent model.
\subsubsection*{\textbf{Quasi particle model ($m(\mu)$)}} 
In the quasi-particle model, the quark mass is assumed to depend on the chemical potential, while the bag pressure is kept constant ($B_0$).
In the dense system, quarks interact with other quarks to create an effective mass, which makes them  behave as quasiparticles. In the hard dense loop approximation, an effective quark propagator generated by resumming one-loop self-energy diagrams is used to determine the zero momentum limit of the dispersion relations, which leads to the effective quark masses.  
\begin{equation}
m^*_f=\frac{m_{f}}{2}+\sqrt{\frac{m_{f}^2}{4}+\frac{g_f^2}{6\pi^2}\mu_f^2}
\end{equation} 

\begin{equation}
    g_f=g_{0}e^{-\alpha_{\mu} \frac{\mu_f}{\mu_0}}
\end{equation} 
Here $\alpha_{\mu}$ is the parameter determining the $\mu$ dependent effective running coupling constant and the value of $\alpha_{\mu}$ should be greater than zero for the restoration of the chiral symmetry. 
The model parameters  $g_0,\alpha_{\mu}, B_0$ are adjusted by the stability criteria \cite{pal:23prd2_bag}.

The thermodynamic potential density for the strange quark matter within
quasiparticle model is as follows
\begin{equation} \label{eq:omega_V_quasi}
\begin{aligned}
 \Omega=-\frac{1}{\pi^2} \sum_{f=u,d,s} \int_0^{k_{f}} \frac{k^4}{\sqrt{k^2 + m_f^{*2}}} \, dk-\frac{1}{2}m_V^2V_0^2+B_0+\Omega_{e}
\end{aligned}
\end{equation} 
The quasiparticle model's $\Omega$ expression \eqref{eq:omega_V_quasi} resembles that of the MIT bag model \eqref{eq:omega_mit_vbag} with the substitution of effective mass for mass and $B_0$ for $B(\mu)$.
Here $B_0$ is a parameter representing the negative vacuum pressure term associated with nonperturbative confinement in QCD. Its introduction allows for the incorporation of confinement effects into theoretical models, and its value is often treated as a free input parameter that can be adjusted or constrained based on physical considerations. The second term represents the thermodynamic potential due to the vector interactions.
The influence of the vector interaction is accounted for by considering the Fermi momentum as :
\begin{equation}
    k_f=\sqrt{(\mu_f-g_VV_0)^2-m_f^{*2}}
\end{equation} 
In a grand-canonical ensemble, the chemical potential in terms of the quark fermi momentum is 
\begin{equation}
\mu_f=\sqrt{k_f^2+m_f^{*2}}+g_VV_0
\end{equation} 

The quark number densities are given by
\begin{equation} \label{eq:quasi_density_zero}
     \rho_f=\frac{k_f^3}{\pi^2}- m_f^{*}\frac{\partial m_f^*}{\partial \mu_f}\frac{3}{\pi^2}\int_0^{k_f}\frac{k^2}{\sqrt{k^2+(m_f^*)^2}}dk 
\end{equation}
The pressure and energy density is obtain using Eq.~\eqref{eq:eos_mit_b_mu}.
\subsubsection*{\textbf{Quark mass density-dependent model ($m(\rho)$)}} 
In this model, medium effects are incorporated through the density-dependent mass. The interactions and medium effects are taken through quark mass $(m_f^*)$. We have taken the quark mass as :
\begin{equation}
m_f^{*}=m_{f}+\frac{C}{\rho^{a/3}}
\end{equation}  
where $a$ and $C$ are model parameters determined by the thermodynamic stability criteria.
The energy density for quark matter with density-dependent quark mass and vector interaction is given by
\begin{equation} 
  \varepsilon=\sum_{f=u,d,s}\frac{3}{\pi^2}\int_0^{k_f} \sqrt{k^2+m_f^{*2}}k^2dk+\frac{1}{2}m_V^2V_0^2+\frac{1}{\pi^2}\int_0^{k_e} k^2\sqrt{k^2+m_e^2}dk
\end{equation} 
The vector field equation reads as
\begin{equation}
    \frac{\partial \varepsilon}{\partial \rho_f} = 0 \implies g_V V_0 = \left( \frac{g_V}{m_V} \right)^2 \sum_{f=u,d,s} \rho_f
\end{equation}
The quark chemical potential is modified due to the density-dependent quark mass and vector interactions.
\begin{equation}
    \mu_f=\frac{\partial \varepsilon}{\partial \rho_f}=\sqrt{k_f^2+m_f^2}+g_VV_0+\frac{3}{\pi^2}m_f^{*}\frac{\partial m_f^*}{\partial \rho_f}\int_0^{k_f}\frac{k^2}{\sqrt{k^2+m_f^{*2}}}
\end{equation}
Pressure is obtained from the Euler relation as expressed in Eq.~\eqref{eq:pressure_euler}.

\section{Results } 
\label{sec:results}

In this work, our primary focus is the study of strange quark stars within the framework of modified $f(R,T)$ gravity. As outlined in the formalism section, we explore two phenomenological quark models while maintaining a self-consistent thermodynamic approach. In our analysis, we have used quark masses of $m_u = 2.0 \, \text{MeV}$, $m_d = 4.67 \, \text{MeV}$, and $m_s = 93.4 \, \text{MeV}$.
For the MIT bag model, we consider four scenarios: (i) a constant bag pressure,(ii)  constant bag pressure with vector interactions,  (ii)  density-dependent bag pressure $(B(\rho))$, with vector interaction and (iii) chemical potential-dependent bag pressure  $(B(\mu))$ with vector interactions. Additionally, in the quark mass model, we examine (i) density-dependent quark mass $m(\rho)$ and (ii)  chemical potential-dependent quark mass $m(\mu)$ with vector interactions.

\begin{figure*}[htp]
\centering
\includegraphics[width=1.0\textwidth]{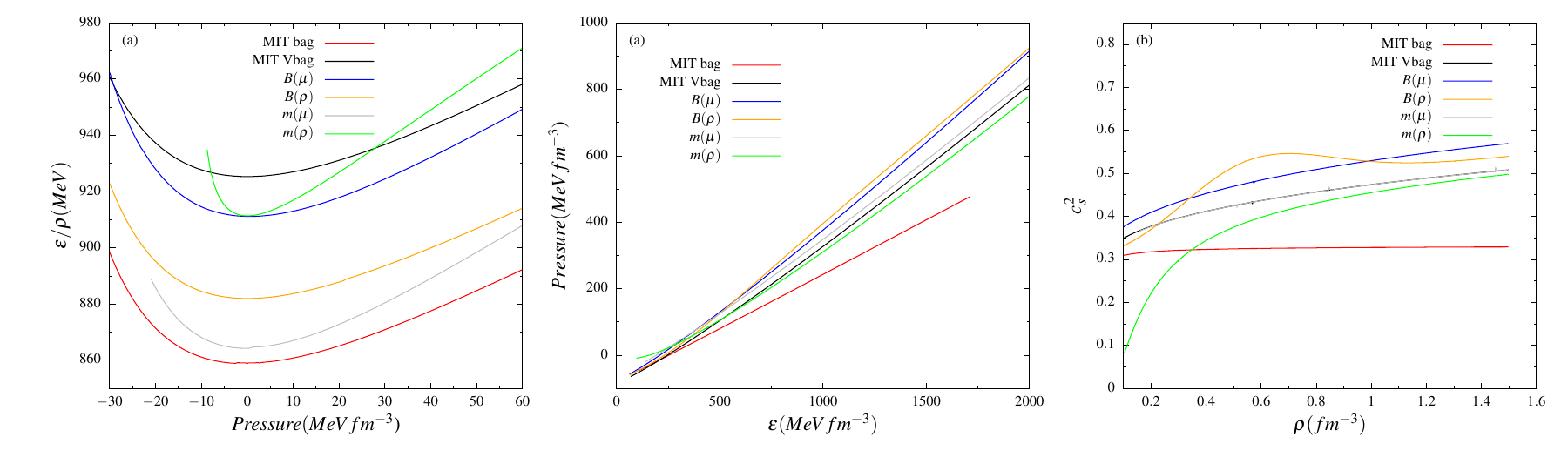}
\caption{The equations of state of different quark models. (a)  variation of energy density per baryon with pressure (b)  variation of pressure with the energy density and (c) speed of sound with baryon density.  }
\label{fig:eos_quark_models}  
\end{figure*}

In Fig.~\ref{fig:eos_quark_models}(a), we examine the thermodynamic stability condition for
the strange quark matter for the six equations of state. From the thermodynamical perspective, the minimum of $\frac{\varepsilon}{\rho}$ should occur exactly at  zero pressure.  In Fig.~\ref{fig:eos_quark_models}(b), we explore the behavior of the equations of state. The inclusion of the vector interaction results in a stiffer equation of state. In Fig.~\ref{fig:eos_quark_models}(c), we show the variations of the speed of sound ($C_s^2$) with density. The incorporation of medium effects causes $C_s^2$ to vary with density. At high densities, it saturates within the range of 0.5 to 0.6. In the absence of vector interactions, its value is approximately equal to one-third.
\begin{figure*}[htp]
\centering
\includegraphics[width=1.0\textwidth]{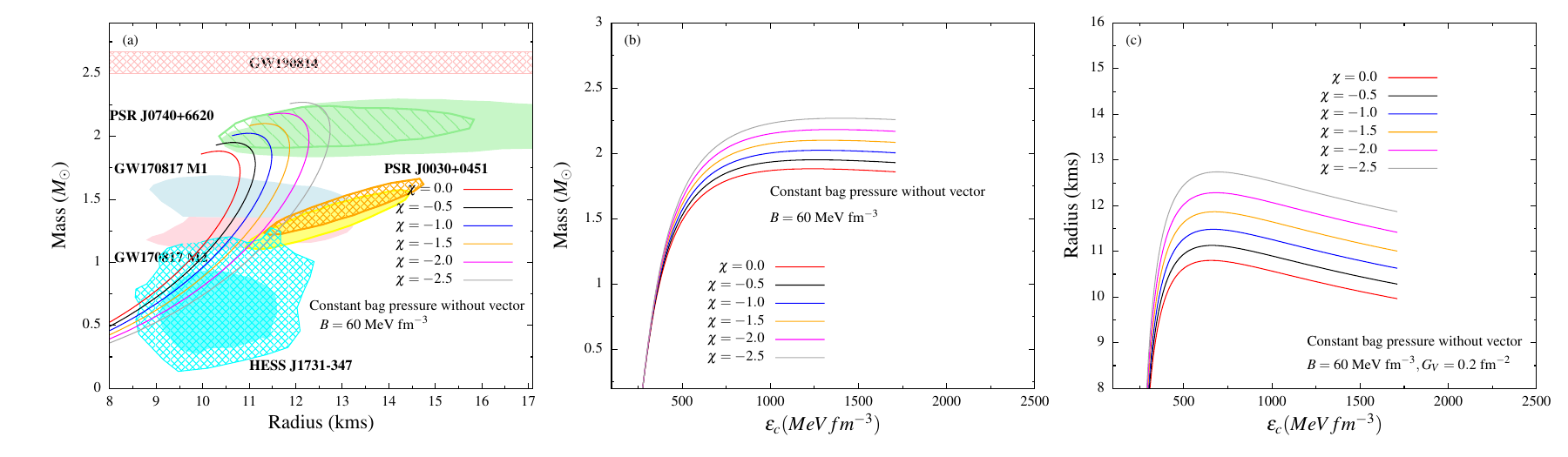}
\caption{ (a)Mass-radius diagram corresponding to the different values of $\chi$ in the MIT bag model without medium effect and vector interaction case (MIT bag). The restrictions on the M-R plane from GW170817 \cite{LIGOScientific:2018cki}, GW190814,  the NICER experiment for PSR J0030+0451 \cite{Riley:2019yda,Miller:2019cac}, PSR J0740+6620\cite{Fonseca:2021wxt}, and HESSJ1731-347 \cite{2022NatAs} have been incorporated. 
(b)The total mass of the star as a function of the central energy density for different values
of $\chi$.
(c)The total radius of the star against the central energy density for some values of $\chi$
 }
\label{fig:mr_bag_const}  
\end{figure*}

We have generated mass-radius curves for all sets of parameterizations and for various values of $\chi$. When $\chi=0$, these curves represent the original TOV mass-radius curve. 

In Fig.~\ref{fig:mr_bag_const}(a), we explore the influence of $\chi$ on the mass-radius diagram within the MIT bag model, assuming a constant bag pressure (MIT bag ). Here, the bag pressure is set to $60~\text{MeV} \text{fm}^{-3}$. It is difficult to satisfy the two solar mass constraints using the constant bag pressure equation of state. The bag pressure is constrained by the Bodmer-Witten conjecture, so we cannot choose arbitrary values for it. This constraint limits the possibility of achieving  $2M_{\odot}$ mass. Alternatively, by using a modified TOV equation \eqref{eq:modi_tov_final}, we can obtain a $2M_{\odot}$ mass with various values of $\chi$ without violating the Bodmer-Witten conjecture. The values of $\chi$ have a significant impact on the mass-radius diagram. As we decrease the values of $\chi$ below zero (negative), the mass-radius curve shifts towards a higher maximum radius and mass. The case of $\chi = 0$ fails to meet the observational constraints from PSR J0030+0451 and PSR J0740+6620. However, with  inclusion of $\chi=-0.5$ onwards, the results satisfy the PSRJ0740+6620 observational constraints. For  value of $\chi$ less than $-2$, the calculations satisfy the  PSR J0030+0451  observational constraints. However none of the $\chi$ values obey the high mass  GW190814 observations as  is observed from the figure.

\begin{figure*}[htp]
\centering
\includegraphics[width=1.0\textwidth]{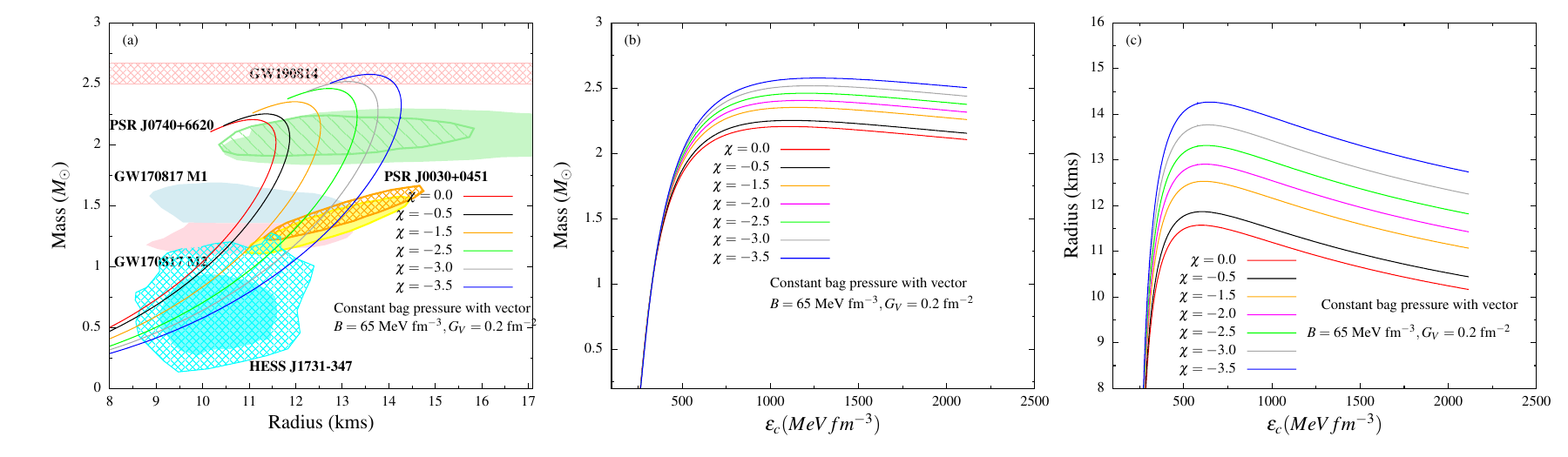} 
\caption{Same as in Fig.~\ref{fig:mr_bag_const} with the MIT bag model with vector interactions (VMIT bag). 
}
\label{fig:mr_bag_const_vector}  
\end{figure*}  

In Fig.~\ref{fig:mr_bag_const_vector}(a), we examine the impact of $\chi$ in the vector MIT bag model with a constant bag pressure (VMIT bag) in the mass-radius diagram. We employed the parameters $B=65 ~\text{MeV}fm^{-3}$, along with a vector interaction parameter of $G_V=0.2 \text{fm}^{-2}$. As in the previous case, without
vector interactions, it is challenging to achieve a $2M_{\odot}$ mass for $\chi=0$. However, this can be attained by incorporating vector interactions, though it still fails to explain the constraints from PSR J0030+0451.  However  inclusion of $\chi$ lesser than $-1.5$ in the equation of state fulfils the  PSR J0030+0451 observational constraints. The value of $\chi$ less than $-3.0$ satisfy the GW190814 observations of  massive compact stars.

\begin{figure*}[htp]
\centering
\includegraphics[width=1.0\textwidth]{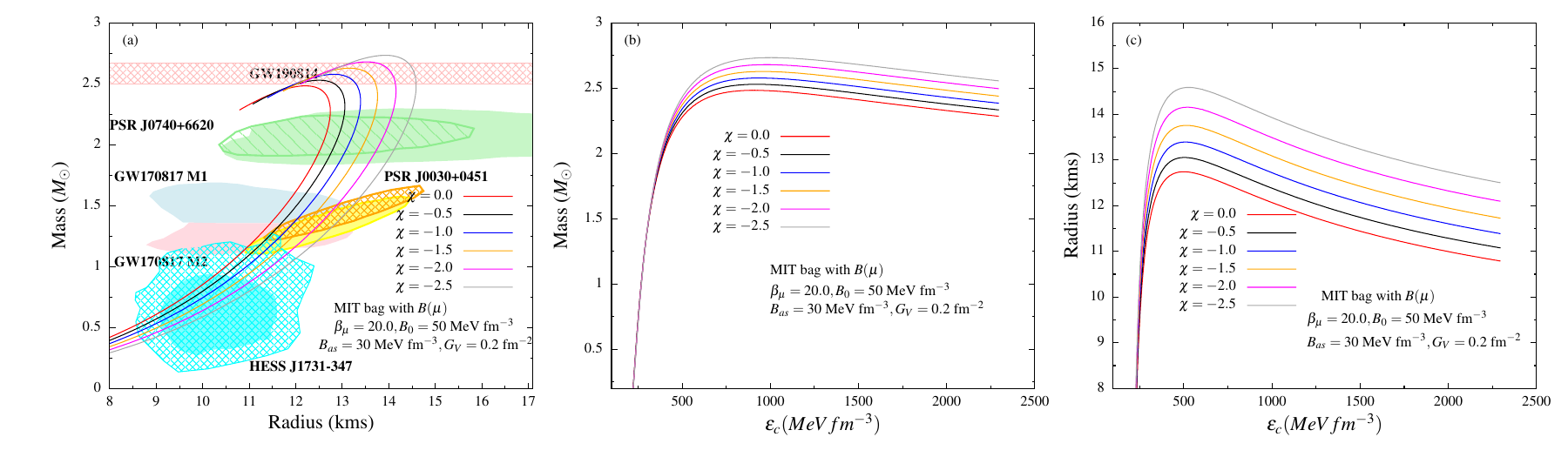} 
\caption{Same as in Fig.~\ref{fig:mr_bag_const} with the chemical potential dependent bag pressure with vector interactions ($B(\mu)$). 
}
\label{fig:mr_b_mu}  
\end{figure*}  

In Fig.~\ref{fig:mr_b_mu}(a), we present the mass-radius diagram, examining the medium effects of the  MIT bag model with the chemical potential-dependent bag pressure ($B(\mu)$). We utilized the parameters $\beta_{\mu}=20$, $B_0=50 \, \text{MeV} \, \text{fm}^{-3}$, $B_{as}=30 \, \text{MeV} \, \text{fm}^{-3}$, and a vector interaction parameter $G_V=0.2 \, \text{fm}^{-2}$. For $\chi=0$, the results are consistent with the observational constraints from HESS J1731-347, GW170817, and PSR J0740+6620. However, for $\chi = 0$, they marginally satisfy the  PSR J0030+0451 and GW190814  observations. When $\chi$ is included, the constraints from PSR J0030+0451 and GW190814 are fully  satisfied.
\begin{figure*}[htp]
\centering
\includegraphics[width=1.0\textwidth]{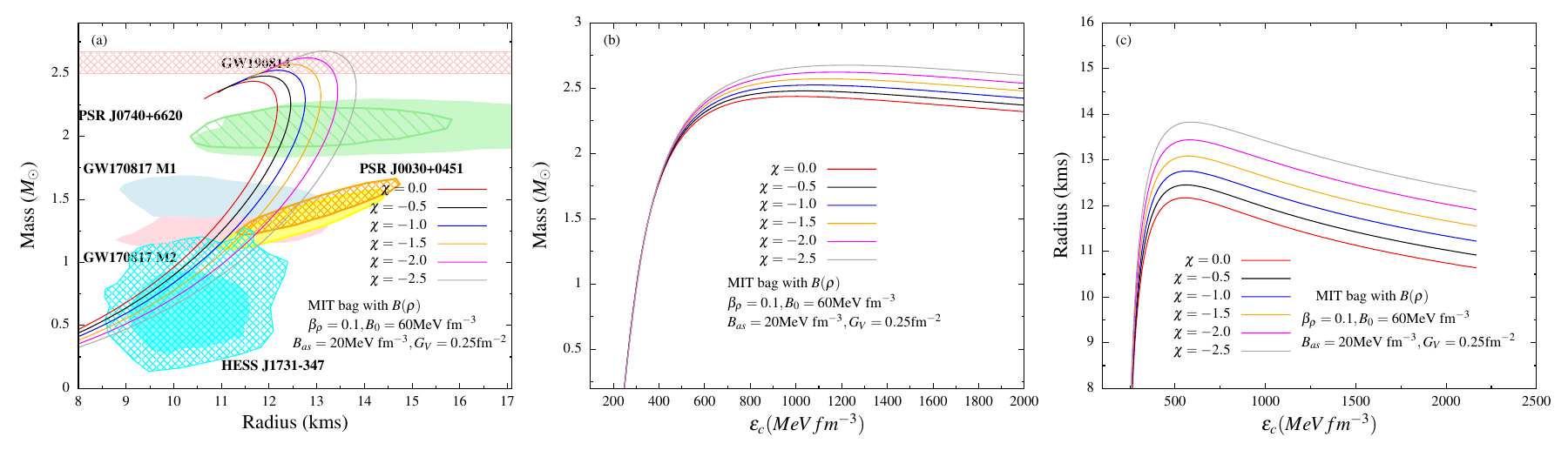} 
\caption{ Same as in Fig.~\ref{fig:mr_bag_const} with the MIT bag model with density-dependent bag pressure and vector interactions ($B(\rho)$). 
}
\label{fig:mr_b_rho}  
\end{figure*}

In Fig.~\ref{fig:mr_b_rho}(a), we present the mass-radius diagram, examining the medium effects of the  MIT bag model with density-dependent bag pressure ($B(\rho)$). We utilized the parameters $\beta_{\rho}=0.1$, $B_0=60 \, \text{MeV} \, \text{fm}^{-3}$, $B_{as}=20 \, \text{MeV} \, \text{fm}^{-3}$, and a vector interaction parameter $G_V=0.25 \, \text{fm}^{-2}$. For $\chi=0$, the results satisfy the observational constraints from HESS J1731-347, GW170817, and PSR J0030+0451. However, these results fail to match the observations of PSR J0030+0451 and GW190814. For $\chi=-1.5$ onwards the results satisfy the PSRJ0030+0451 observations. GW190814 constraints are satisfied for $\chi$ values less than $-0.5$.

\begin{figure*}[htp]
\centering
\includegraphics[width=1.0\textwidth]{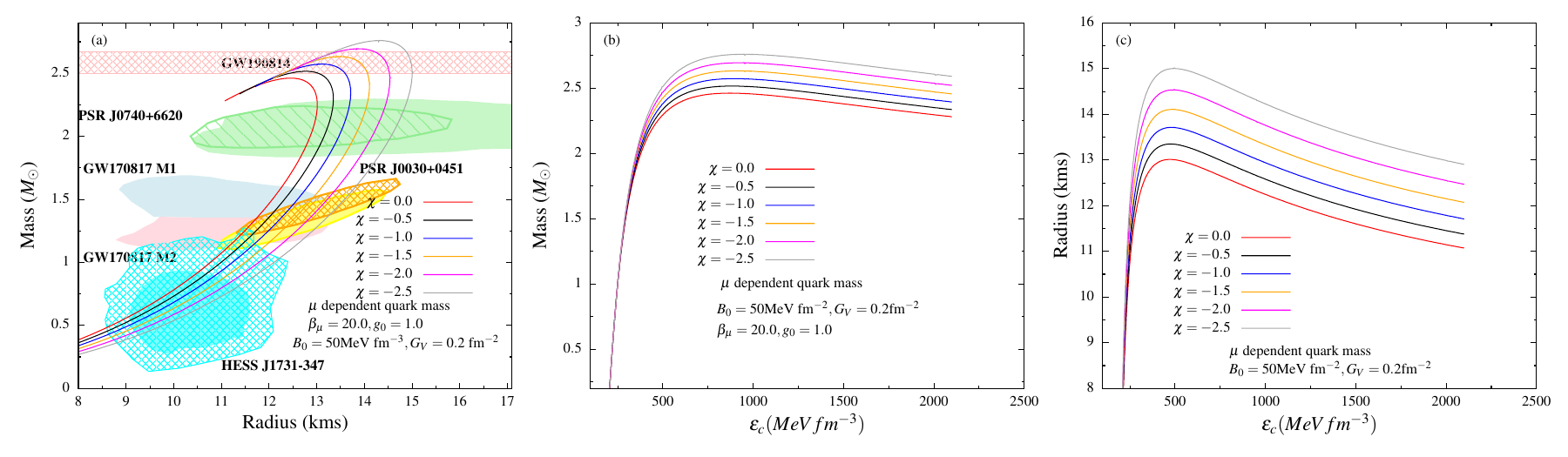} 
\caption{ Same as in Fig.~\ref{fig:mr_bag_const} with the quark mass model with the chemical potential-dependent quark mass model with vector interactions ($m(\mu)$). 
}
\label{fig:mr_m_mu}  
\end{figure*} 
In Fig.~\ref{fig:mr_m_mu}(a), we present the mass-radius diagram, examining the medium effects of the quark mass model with the chemical potential-dependent quark mass ($m(\mu)$). We utilized the parameters $\alpha_{\mu}=20$, $g_0=1.0$, $B_{0}=50 \, \text{MeV} \, \text{fm}^{-3}$, and a vector interaction parameter $G_V=0.2 \, \text{fm}^{-2}$. For $\chi=0$, the results are consistent with the observational constraints from HESS J1731-347, GW170817, PSR J0740+6620, and partially PSR J0030+0451. However, for $\chi = 0$, the findings do not match the observations from  GW190814. When $\chi$ is included, the constraints from PSR J0030+0451 and GW190814 are successfully satisfied.

\begin{figure*}[htp]
\centering
\includegraphics[width=1.0\textwidth]{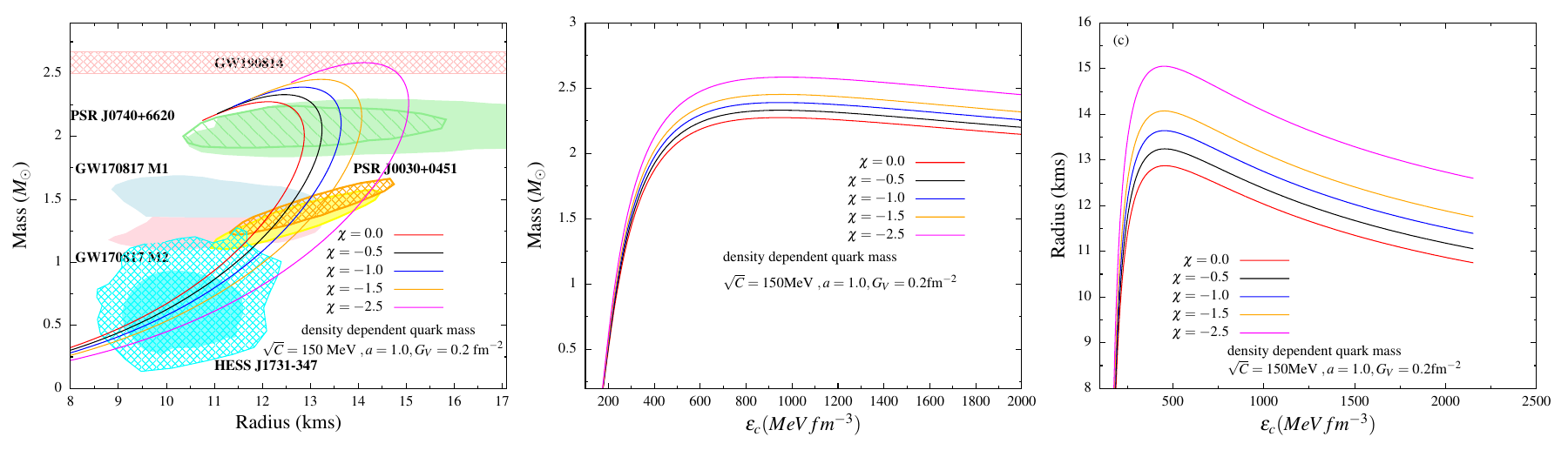} 
\caption{ Same as in Fig.~\ref{fig:mr_bag_const} with the the density-dependent quark mass mass model with vector interactions ($m(\rho)$). 
}
\label{fig:mr_m_rho}  
\end{figure*}   
In Fig.~\ref{fig:mr_m_rho}(a), we present the mass-radius diagram, examining the medium effects of the quark mass model with density-dependent quark mass ($m(\rho)$). We utilized the parameters $\sqrt{C}=150 \,\text{MeV} \,$, $a=1.0$, and a vector interaction parameter $G_V=0.2 \, \text{fm}^{-2}$. For $\chi=0$, the results are consistent with the observational constraints from HESS J1731-347, GW170817, PSR J0740+6620, and  PSR J0030+0451. However, for $\chi = 0$, the findings do not match the observations from  GW190814. When $\chi$ is included, the constraints from  GW190814 are successfully satisfied.

The behavior of the total mass is displayed in Fig.~\ref{fig:mr_bag_const} --\ref{fig:mr_m_rho} (b) against the central energy density( $\varepsilon_c$ ) for the strange quark stars using the above mentioned six equations of state for different $\chi$ values. It is observed that at lower values of the central energy density, the effect of $\chi$ is minimal. The variation of the total mass with  central energy density is quite similar though their magnitude differs slightly as seen from the figures. The maximum mass is reached at around the central energy  density ($\varepsilon_c$) range of $(400-600)~ \text{MeV fm}^{-3}$ after which it saturates or slightly decreases at higher central energy density values. The maximum mass attained however varies with the equations of state as well as the magnitude of $\chi$ used.

In ~~Fig.~\ref{fig:mr_bag_const} --\ref{fig:mr_m_rho}(c), the behavior of the total radius is shown against the central energy density $\varepsilon_c$ for the strange quark stars for the different equations of state. As seen in the Figs.~\ref{fig:mr_bag_const} --\ref{fig:mr_m_rho}(b), it is similarly observed that at  lower values of the central energy density, the  the total radius of the stars is insensitive to the value of $\chi$.
The  pattern of variation of the total radius with  central energy density is quite similar for different equations of state though their magnitude differs slightly as is seen from the figures.
At the lower central energy density the values of the total radius increases sharply and then  reaches a peak around the central energy  density ($\varepsilon_c$) range of $(400-600)~ \text{MeV fm}^{-3}$ after which it gradually decreases at higher central energy density range. This is in contrast to the variation of the maximum mass which more or less saturates at higher energy densities.  The radius corresponding to the maximum mass varies with the equations of state as well as the magnitude of $\chi$ used.

\begin{figure*}[htp]
    \centering
        \includegraphics[width=1.0\textwidth]{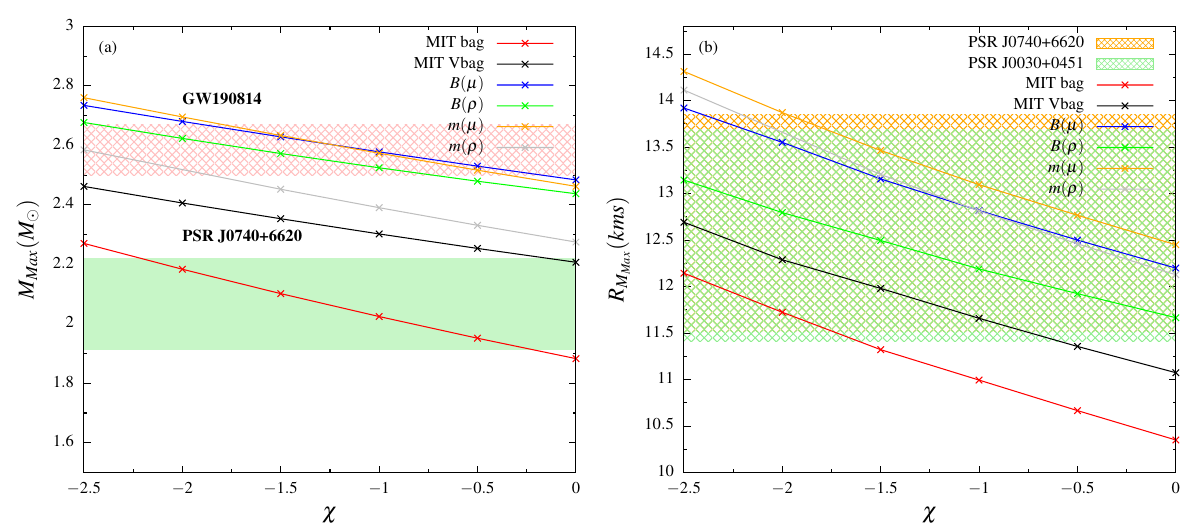} 
\caption{The impact of $\chi$ on (a) the maximum mass of quark stars and (b) their corresponding radius. }
\label{fig:MR_max_chi}  
\end{figure*}

In Fig.~\ref{fig:MR_max_chi}, we summarise the effects of $\chi$ on the maximum mass ($M_{Max}$) of the strange quark star and the  the radius ($R_{M_{Max}}$) corresponding to the ($M_{Max}$) for the six quark equations of state.  It is observed that there has been a considerable increase  in  values of both $M_{Max}$ and $R_{M_{Max}}$ as one includes $\chi$ and changes its value for all the equations of state.
It is seen that both ($M_{Max}$) and ($R_{M_{Max}}$) scales with $\chi$ in similar fashion for the different equations of state. The fitted relations for $M_{Max}$ and $R_{M_{Max}}$  are as follows:
\begin{equation}
    \begin{aligned}
        M_{Max}&=a_M+b_M\chi,~\\
        R_{M_{Max}}&=a_R+b_R\chi+c_R \chi^2.~\\
    \end{aligned}
\end{equation}
where $a_M,b_M,a_R,b_R,c_R$ are the fitted parameters. These values are different for different equations of state as shown in Tab.~\ref{tab:fitted_params}.

\begin{table}[!ht]
\centering
\caption{Fitted parameters for $M_{\text{Max}}$ and $R_{\text{Max}}$ as functions of $\chi$ for six equations of state.}
\setlength{\tabcolsep}{10 pt}
\begin{tabular}{cccccc}
\hline
\hline
Model & $a_M$ & $b_M$ & $a_R$ & $b_R$ & $c_R$ \\
\hline
MIT bag & -0.154749  &  1.87541   &   0.05875 &  -0.565902   & 10.3578\\ 
MIT Vbag  & -0.10192   &  2.20303   &   0.0449829   &  -0.528175   &  11.0783 \\
$B(\mu)$& -0.100257  &  2.4805    &   0.0459929   &    -0.576549 &     12.1986 \\
$B(\rho)$& -0.0958589 &  2.43229   &   0.0421786   &  -0.484554   &     11.6672 \\
$m(\mu)$ & -0.119246  &  2.45766  &    0.06705     &   -0.575741  &     12.4538 \\
$m(\rho)$ & -0.124304 &  2.26991  &    0.0727228   &  -0.61111    &     12.1306 \\
\hline
\hline
\end{tabular}
\label{tab:fitted_params}
\end{table}

In Fig.~\ref{fig:MR_max_chi}(a), we find that the maximum mass for the equation of state based on the MIT bag model without vector interactions, satisfy the observations of PSR J0740+6620 \cite{Riley:2021pdl}, except for the cases $\chi=0$ and $\chi=-2.5$. Additionally, these $\chi$ values fail to satisfy the GW190814 observations\cite{LIGOScientific:2020zkf}. In the case of the MIT bag model with vector interactions ( without medium effcets), only the maximum mass for $\chi=0$ satisfy the constraints from PSR J0740+6620 \cite{Riley:2021pdl}. For other models (with medium effects ), the maximum mass does not satisfy the the PSRJ0740+6620 observations. In the case of the density-dependent quark mass model ($m(\rho)$), only the $\chi=-2.5$ case satisfies the maximum mass constraints from GW190814 \cite{LIGOScientific:2020zkf}. In the case of the $\mu$-dependent bag model ($B(\mu)$) and the $\mu$-dependent quark mass model ($m(\mu)$), the maximum mass constraints from GW190814 are satisfied for $\chi$ in the range $[-0.5, 2.0]$. However, for the $\rho$-dependent bag model ($B(\rho)$), the cases $\chi=0$ and $\chi=-0.5$ are excluded from the observational data range, remaining values of  $\chi$ satisfies the GW190814 observations.

In Fig.~\ref{fig:MR_max_chi}(b), we imposed radius constraints from the PSR J0030+0451 \cite{Riley:2019yda} and PSR J0740+6620 \cite{Riley:2021pdl} observations. We find that for the MIT bag model without vector interactions, the range $\chi > -1.5$ is excluded,  while for the model with vector interactions, $\chi > -0.5$ is excluded by the observational constraints. In the $\mu$-dependent bag model ($B(\mu)$) and the quark mass models ($m(\mu)~\text{and}~m(\rho) $), $\chi = -2.5$ does not satisfy the radius constraints. In the density-dependent bag model ($B(\rho)$), $\chi$ values satisfy the radius constraints.

In the MIT bag model, the value of $\chi=-2$ satisfies  both   the mass and radius constraints better as seen from  the Figs.~\ref{fig:MR_max_chi}(a) and \ref{fig:MR_max_chi}(b). Amongst the other five equations of state, density dependent bag model ($B(\rho)$) in the framework of modified gravity  is found to be most suitable for satisfying the constraints of both the mass and radius as seen from  the Figs.~\ref{fig:MR_max_chi}(a) and \ref{fig:MR_max_chi}(b).
\begin{figure*}[htp]
    \centering
        \includegraphics[width=1.0\textwidth]{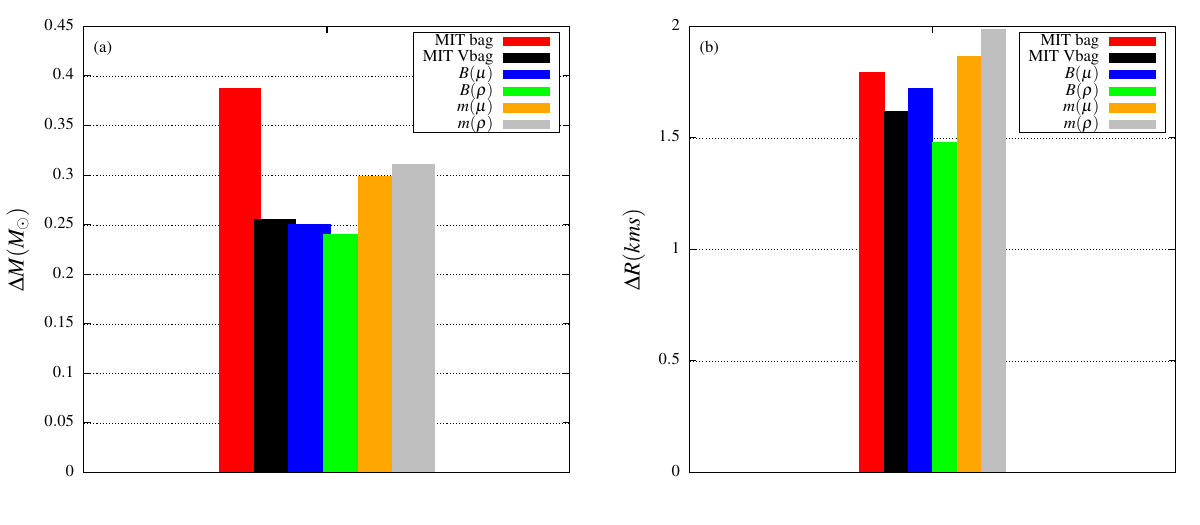} 
\caption{(a)The maximum mass difference and (b) corresponding radius differences for the six equation of state }
\label{fig:deltamr}  
\end{figure*} 

It is observed from the  Fig.~\ref{fig:deltamr}(a) that the maximum mass $M_{Max}$ is increased by $( 0.23-0.27)$ $M_{\odot}$ and 
the corresponding change in the radius $R_{M_{Max}}$ is  more or less  ($1.5-2.0)~\text{kms}$ as seen from the Fig.~\ref{fig:deltamr}(b) when $\chi$ changes from $0$ to $-2.5$ for the different EoS used. The maximum mass is increased most in the MIT bag model( without vector interaction and medium effects) as seen from Fig.~\ref{fig:deltamr}(a). The change in the radius ($\Delta R$) is maximum in the case of the density-dependent quark mass model ($m(\rho)$).
\begin{figure*}[htp]
    \centering
        \includegraphics[width=0.45\textwidth]{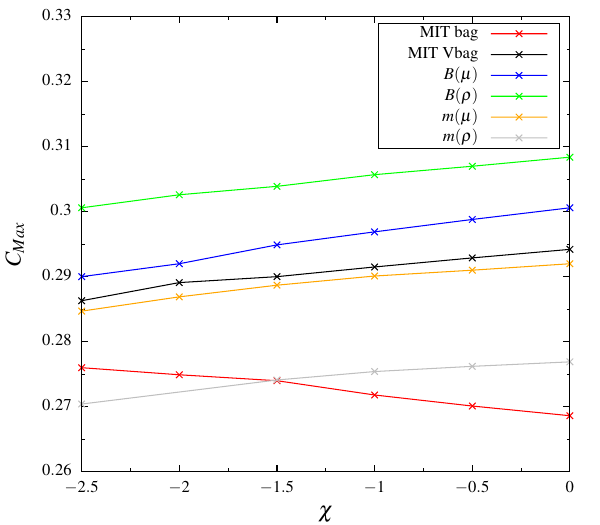} 
\caption{The dependence of compactness parameter $C_{Max}$ with $\chi$. }
\label{fig:c_max_chi}  
\end{figure*} 
In Fig.~\ref{fig:c_max_chi}, we study the compactness parameter 
$C_{\text{Max}} = \frac{M_{\text{Max}}}{R_{M_{\text{Max}}}}$
as a function of $\chi$. $C_{\text{Max}}$  increases with $\chi$ for the vector interaction cases, while for the case without vector interaction (simple MIT bag model), \(C_{\text{Max}}\) decreases with $\chi$.

\section{Summary and conclusion } 
\label{sec:conclusion} 
In this work, we have considered the modified theory of gravity and consequently have solved the modified TOV equation for the calculation of the mass-radius of quark stars. We have taken the six quark equations of state by using some variants of the MIT bag model and the quark mass model. All the model parameters in the quark equations of state are chosen by respecting the Bodmer-Witten conjecture. We examined the self-consistent thermodynamics of the equations of state and observed that the minimum energy density per baryon density is achieved precisely at zero pressure.
 We also found that  using the negative values of $\chi$ has increased the mass and radius of the quark stars. We have also seen that  it is difficult to satisfy the PSR J0030+0451 constraints using the MIT bag model without including   the 
 vector interaction. However, with  the inclusion of  $\chi$, the results satisfied the observational constraints. In the other equations of state, the modified theory of gravity aids in satisfying the constraints imposed by the GW190814 event. It is noteworthy that the inclusion of $\chi$ allows us to address the high mass constraints from GW190814 and also NICER PSR J0030+0451. Additionally, we provided a constraint on $\chi$ for each equation of state, based on the maximum mass and corresponding radius derived from the observations.

\bibliography{ref}  
\bibliographystyle{JHEP}
\end{document}